# Strain-Induced Boundary States and Phase Transitions in Graphene Flakes

Yongsheng Liang[1], Shiqi Xia[1*], Daohong Song[1,2], Hrvoje Buljan[1,3*], Zhigang Chen[1,2*]

[1]*The MOE Key Laboratory of Weak-Light Nonlinear Photonics, TEDA Institute of Applied Physics and School of Physics, Nankai University, Tianjin 300457, China*

[2] *Collaborative Innovation Center of Extreme Optics, Shanxi University, Taiyuan, Shanxi 030006, China*

[3] *Department of Physics, Faculty of Science, University of Zagreb, Zagreb, Croatia*

* shiqixia@nankai.edu.cn, hbuljan@phy.hr, zgchen@nankai.edu.cn

**Abstract:**

Strain has been extensively employed to tailor graphene's properties and has emerged as a powerful tool for engineering gauge fields and exploring fundamental phenomena in artificial platforms like photonic graphene. Here we discover that, in graphene flakes with custom boundaries, one can create or destroy edge states depending on the direction of the applied uniaxial strain. This is experimentally demonstrated in a photonic platform with two specific examples: one flake structure with pairs of twig and zigzag edges, and the other with pairs of armchair and bearded edges. We find that the existence of the edge states and their positions in momentum space are accurately predicted with appropriate winding numbers, unveiling the underlying topology of such edge states. Furthermore, when a graphene flake supports the maximum number of edge states along boundaries after a semimetal-to-insulator transition, both compact localized edge and corner states emerge, indicating the realization of a photonic minimal-model higher-order topological insulator based on such strained graphene flakes.

# 1 Introduction

Graphene, a two-dimensional material with a honeycomb lattice (HCL) structure, has emerged as a cornerstone material in modern condensed matter physics, thanks to its extraordinary combination of electronic, optical, and mechanical properties [1-3]. Of particular interest are graphene ribbons and flakes, whose properties are strongly affected by their edges [4-8]. Ribbons with zigzag edges, for example, host zero-energy localized states, whereas those with armchair edges do not support any edge states (within the tight-binding model) [5, 9]. Over the decades, numerous efforts have focused on addressing intriguing questions mediated by graphene edges. For example, how would edge defects and active edge tailoring affect the edge states? What potential applications would edge effects have in the development of electronic and photonic devices? More recently, strain engineering has emerged as a powerful technique for manipulating the physical properties of graphene [10-12]. For instance, by introducing strain, it is possible to shift the Dirac cones, open the bandgap, and even alter its electronic and optoelectronic properties to realize zero-field quantum Hall effect [13-16].

Apart from electronic graphene, synthetic HCL platforms have emerged in various realms, including photonics, acoustics, mechanical, and electronic circuits. Such artificial graphene can mimic the wave dynamics and phenomena present in electronic graphene while providing enhanced control over the lattice structures [17], such as precise edge tailoring [18-20]. Synthetic HCLs have facilitated studies of Floquet and valley Hall topological insulators [21-26], as well as edge states in a ribbon with bearded or twig boundaries [18, 19, 27]. Furthermore, strain engineering—using uniaxial or complex strain—has been realized in synthetic HCLs, inducing artificial gauge fields and creating Landau levels [28-32]. Uniaxial strain in HCLs can lead to the merging of the Dirac cones and a semimetal-to-insulator phase transition [33-35], and it can also induce the creation of edge states under specific boundary conditions [36, 37]. However, strain engineering of graphene flakes with specially tailored boundaries is largely unexplored, especially regarding the underlying topological features. Many intriguing questions remain. For example: Can boundary states in custom-designed graphene flakes be manipulated through direction-dependent strain? Can topological phases and phase transitions be uncovered in such systems?

In this work, we explore the interplay between strain engineering and boundary tailoring of graphene (HCL) flakes. By using a direction-dependent uniaxial strain, we find that the graphene flake can undergo an unconventional semimetal-to-insulating phase transition. Such a phase transition in

HCL flakes is accompanied by the emergence or suppression of topological boundary states, including both edge and corner states. By systematically analyzing the four fundamental boundary types in graphene (zigzag, armchair, bearded, and twig), we reveal how these edge terminations influence the formation of boundary states and drive phase transitions in graphene flakes. Furthermore, the possibility of interpreting strained graphene in the insulating regime as a minimal model of higher-order topological insulators (HOTIs) is discussed [38, 39], where the emergence of integer bulk polarization and a corner-induced filling anomaly jointly give rise to topological corner states and maximize boundary-state formation. Experimentally, compact edge states originating from the formation of topological flatbands, along with corner states, are observed in a finite-size photonic graphene with custom boundaries, providing evidence for such a distinct phase transition.

## 2 Principles and Methods

### 2.1 Strain-Controlled Boundary States in Graphene Flakes

The HCL comprises two sublattices ($A$ and $B$) within one unit cell, depicted by white and black dots in Fig. 1(b). We consider two HCL flakes with distinct boundary conditions along perpendicular directions: (i) a twig-zigzag flake with pairs of twig and zigzag boundaries (Fig. 1(a)), and (ii) an armchair-bearded flake with pairs of armchair and bearded boundaries (Fig. 1(c)). Two types of unit cells (green and purple rhombuses in Fig. 1(b)) are selected to describe the physical edge properties of the associated HCL flakes. The green (purple)-shaded unit cell corresponds to the twig-zigzag (armchair-bearded) flake, matching the twig and zigzag (armchair and bearded) boundaries. In the tight-binding model, the three nearest-neighbor couplings are denoted as $t_1$, $t_2$, and $t_3$. The bulk Bloch Hamiltonians for the two types of unit cells in Fig. 1(b) can be expressed as:

$$H(\mathbf{k}) = \begin{pmatrix} 0 & h(\mathbf{k}) \\ h^*(\mathbf{k}) & 0 \end{pmatrix} \quad (1)$$

where $h(\mathbf{k}) = h_{TZ}(\mathbf{k}) = t_1 e^{i\mathbf{k}\mathbf{a_2}} + t_2 e^{i\mathbf{k}\mathbf{a_1}} + t_3$ accounts for the green-shaded unit cell, and $h(\mathbf{k}) = h_{AB}(\mathbf{k}) = t_1 + t_2 e^{i\mathbf{k}\mathbf{a_4}} + t_3 e^{i\mathbf{k}\mathbf{a_3}}$ for the purple-shaded unit cell. As shown in Fig. 1(a, c), the basis vectors are given by $\mathbf{a_1} = 1\hat{x}$, $\mathbf{a_2} = ((1\hat{x} - \sqrt{3}\hat{y}))/2$ and $\mathbf{a_3} = ((1\hat{x} - \sqrt{3}\hat{y}))/2$, $\mathbf{a_4} = ((-1\hat{x} - \sqrt{3}\hat{y}))/2$ for the respective unit cells. In the case of an unstrained HCL under periodic boundary conditions, both flakes have the same bulk band structure, with Dirac band touching points located at the corners of the first Brillouin zone (BZ). These degenerate topological singularities characterize a topological semimetal regime [40]. The existence of topological edge states is related to the position of the Dirac points, and is precisely predicted by the nontrivial winding number:

$$w = \frac{1}{2\pi} \oint \frac{d}{d\mathbf{k}} \arg[h(\mathbf{k})] \, d\mathbf{k}, \qquad (2)$$

where $h(\mathbf{k})$ is the off-diagonal term of the bulk Hamiltonian $H(\mathbf{k})$ in Eq. (1). In an unstrained flake, a flatband of edge states is supported by the twig boundary, but a mini-flatband (existing only in a limited BZ regime) is supported by the zigzag/bearded boundary. More details can be found in Supplementary Materials (SM) [41].

By introducing uniaxial strain (i.e., compression for the case studied here) along a specific coupling direction, a shift of the Dirac points is induced along the edges of the BZ, perpendicular to the strain direction (Fig. 1(b)). This shift may alter the existence region of edge states in momentum space and, consequently, affect the number of edge states in flakes with custom boundaries. To describe this process, we define the coupling ratio $\delta_n = t_n/t_0$, where $n = 1, 2,$ or $3$ corresponds to the coupling ratio along one of three (individual strain) directions, and $t_n = t_0$ in an unstrained HCL (Fig. 1(b)). As an example, here we consider a strain applied along the $y$-direction (Fig. 1(b)), which includes all essential results. Under this strain, the coupling ratio $\delta_1 = t_1/t_0$ increases and the nonequivalent Dirac points move towards each other along the $k_x$-direction. Although in the semimetal regime ($\delta_1 < 2$), both boundary-obstructed flakes support edge states that are degenerate at zero energy (Fig. 2(a2, b2)), the strain creates edge states in the twig-zigzag flake but destroys edge states in the armchair-bearded flake (blue dots in Fig. 2(a3, b3)). Beyond a transition threshold ($\delta_1 > 2$), the strain opens a full band gap due to the annihilation of Dirac points and drives the system into the insulating regime, in which a dramatic difference is evident in the number of supported edge states. The twig-zigzag flake hosts a complete flatband of edge states at both boundaries, maximizing the number of edge states (Fig. 2(a3)), characterizing the nontrivial insulating phase. In contrast, the edge states vanish in the armchair-bearded flake (Fig. 2(b3)), indicating a trivial insulating phase. Other scenarios where strain is applied along different directions (thus affecting couplings $t_2$ or $t_3$) are discussed further in the SM [41], with the main features schematically illustrated in Fig. 1(a, c). This finding highlights the crucial role of uniaxial strain in manipulating edge states in graphene flakes with custom boundaries.

*2.2 Minimal-Model Higher-Order Topological Insulator*

Furthermore, we explore the possibility of interpreting these systems as the minimal model for HOTIs, which exhibits both similarities and differences with conventional HOTIs, such as polarization, fractional corner charge, and the degeneracy of boundary states. The bulk polarizations of the flakes are shown in Fig. 2(a3, b3), calculated according to the definition:

$$P_m = -\frac{1}{S}\iint_{BZ} A_m d^2\mathbf{k} \tag{3}$$

where $A_m = -i\langle u(\mathbf{k})|\partial_{\mathbf{k}_m}|u(\mathbf{k})\rangle$ is the Berry connection with $m = x, y$, $u(\mathbf{k})$ is the eigenstate of $H(\mathbf{k})$, and $S$ is the area of the first BZ. The integration is carried out over the first BZ. For strain applied along the $y$-direction, after the gap opens, the polarization takes the values of $(\sqrt{3}/4, -3/4)$ and $(0, 0)$ for the bulk Hamiltonian with $h_{TZ}(\mathbf{k})$ and $h_{AB}(\mathbf{k})$, respectively. Consequently, the Wannier centers shift to the edges of the unit cells in the twig-zigzag flake (red markers on green rhombus in Fig. 2(a1, b1)), leading to a fractional charge ($\sigma = 1/2$) along the boundaries, which guarantees the presence of edge states [42]. Additionally, a fractional corner anomaly ($\phi$) can be extracted at $C_2$-symmetric corners (left-top and right-bottom corners in Fig. 2(a1)), given by $\phi = (p - (2\sigma)) \mod 1 = 1/2$, where $p = 1/2$ is the corner charge. The non-zero bulk polarization and fractional corner charge in the nontrivial insulating regime signify the emergence of higher-order topological features and predict the presence of corner states in the strained twig-zigzag flake (red lines in Fig. 2(a2)). In contrast, in the armchair-bearded flake, the Wannier centers lie at the centers of the unit cells (Fig. 2(b1)), thus both edge and corner states are absent in the trivial insulating regime (Fig. 2(b2)). Notice that not all strained graphene flakes in the nontrivial insulating regime exhibit HOTI characteristics. HOTI behavior emerges only when both boundaries support flatband edge states. For example, when a strain is applied along $t_2$ in a twig-zigzag flake, the system does not show the HOTI phase, retaining zero component of the bulk polarization and an integral-valued corner charge, despite being in a nontrivial insulating regime [41].

More interestingly, unlike conventional HOTIs, where the bulk polarization is always quantized and the gap closes only at a specific point in parameter space, a distinctive characteristic of the strained graphene is the phase transition from a semimetal to an insulating regime [33-35]. The topological properties of edge modes in the semimetal regime persist in the nontrivial insulating regime due to preserved chiral symmetry and nontrivial winding in momentum space. As a result, if a topological corner state appears in the nontrivial insulating regime of strained graphene, it exhibits an atypical degeneracy with the zero-energy topological flatband edge states. As shown in Fig. 2(a5), the energy distributions of degenerate corner states and compactly-localized edge states (CESs) are confined to one sublattice at each end - a result of chiral symmetry, allowing energy to be confined at any position along the boundaries in the nontrivial insulating regime. Therefore, the characteristic phase transition and the presence of degenerate topological edge and corner states highlight the differences between strained graphene flakes and conventional HOTIs.

# 3 Results

*3.1 Experimental Observation of Extended and Compact Edge States*

To experimentally demonstrate the evolution of the edge states, photonic graphene with the desired boundary condition is fabricated using the continuous-wave laser-writing technique in a nonlinear crystal (SBN) [43]. The HCL structure under the twig-zigzag boundary condition and the corresponding 1D band structures of zigzag ribbon under different coupling ratios are shown in Fig. 3(a, b). By using a spatial light modulator, a probe beam matching the zigzag edge states at the $\Gamma$ point is generated (Fig. 3(c1, c2)). The output of the light beam resides only on the $A$ sublattices and exponentially decays into the bulk. In the semimetal regime, however, the probe beam fails to preserve its shape and light couples into the $B$ sublattices after $20\ mm$ propagation (Fig. 3(c3)). In contrast, in the nontrivial insulating regime, the probe beam remains localized and intact after propagation through the HCL (Fig. 3(c4)). Similar phenomena are observed under the twig boundary condition (Fig. 3(d)), where the edge states at the $\Gamma$ point are preserved only in the insulating regime (Fig. 3(d4)). The presence of zigzag edge states at the $\Gamma$ point (red star in Fig. 3(b2)) indicates the opening of the bulk band gap and signals the semimetal-to-insulator phase transition. More details can be found in SM [41]. Moreover, the formation of topological flatbands in the nontrivial insulating phase contributes to the presence of CES (Fig. 3(e1)). As a result, the CES is preserved (Fig. 3(e4)) in the insulating regime but deteriorated in the semimetal regime (Fig. 3(e3)). The distribution of edge states at the $\Gamma$ point and the corresponding simulations with longer propagation distances are included in the SM [41].

*3.2 Experimental Observation of Corner States in the HOTI Regime*

To demonstrate the characteristic HOTI feature of strained graphene, corner states are observed in the nontrivial insulating regime of the photonic twig-zigzag flake (Fig. 4(a)). The probe beam (Fig.

4(c1)) is modulated to match the mode distribution of the corner state (Fig. 2(a4)) and launched into the corner sites (dashed red rectangle in Fig. 4(a)). After $20\ mm$ propagation, the output of the probe beam remains localized at the initially excited A sublattice sites (Fig. 4(c2)), with no light coupling to the $B$ sublattices, confirming the presence of corner states. For a direct comparison, a photonic graphene flake with identical spatial parameters but under the armchair-bearded boundary condition is constructed, which is in the trivial insulating regime, i.e., the trivial HOTI phase (Fig. 4(b2)). In this case, the corner states are absent and the probe beam (Fig. 4(d1)) becomes strongly distorted and spreads into the neighboring $B$ sublattices after 20 mm propagation (Fig. 4(d2)). Numerical simulations for a longer propagation distance ($80\ mm$) further highlight the contrast between the nontrivial (Fig. 4(c3)) and trivial (Fig. 4(d3)) insulating regimes. These results, consistent with the theoretical prediction, confirm that the strained twig-zigzag flake under the $y$-directional strain can support both topological edge and corner modes in the nontrivial insulating regime, manifesting the HOTI features. Note that the HOTI corner states observed here are fundamentally different from those previously realized merely by boundary reconfiguration, but without strain engineering [44].

## 4 Discussion and Conclusion

In conclusion, we have shown that uniaxial strain can serve as an effective approach for manipulating boundary states and realizing topological phase transitions in graphene flakes with custom boundaries. Theoretical analysis and experimental demonstration using laser-written photonic graphene confirm that the creation or destruction of boundary states in graphene flakes is closely related to strain direction and boundary conditions. Importantly, we identify a nontrivial insulating regime where the HOTI features emerge, giving rise to corner states under specific strain conditions. Moreover, the semimetal-to-insulator phase transition and the degenerate topological edge and corner states in strained graphene highlight its distinct features that differ from conventional HOTIs. The similarities and differences between strain-induced HOTIs and those previously established HOTIs certainly merit further investigation, especially when the minimal HOTI model is relevant. These results establish strained graphene flakes as versatile platforms for engineering boundary phenomena, which may offer new opportunities for applications in topological lasers and quantum emitters, as well as for exploring fundamental physics in photonics and beyond [45-49].

**Acknowledgments:**

This work was supported by National Key R&D Program of China (No. 2022YFA1404800); the National Nature Science Foundation of China (No. W2541003, 12134006, 12274242, and 12474387); the Natural Science Foundation of Tianjin (No. 21JCJQJC00050) and the 111 Project (No. B23045) in China. H.B. acknowledges support by the QuantiXLie Center of Excellence.

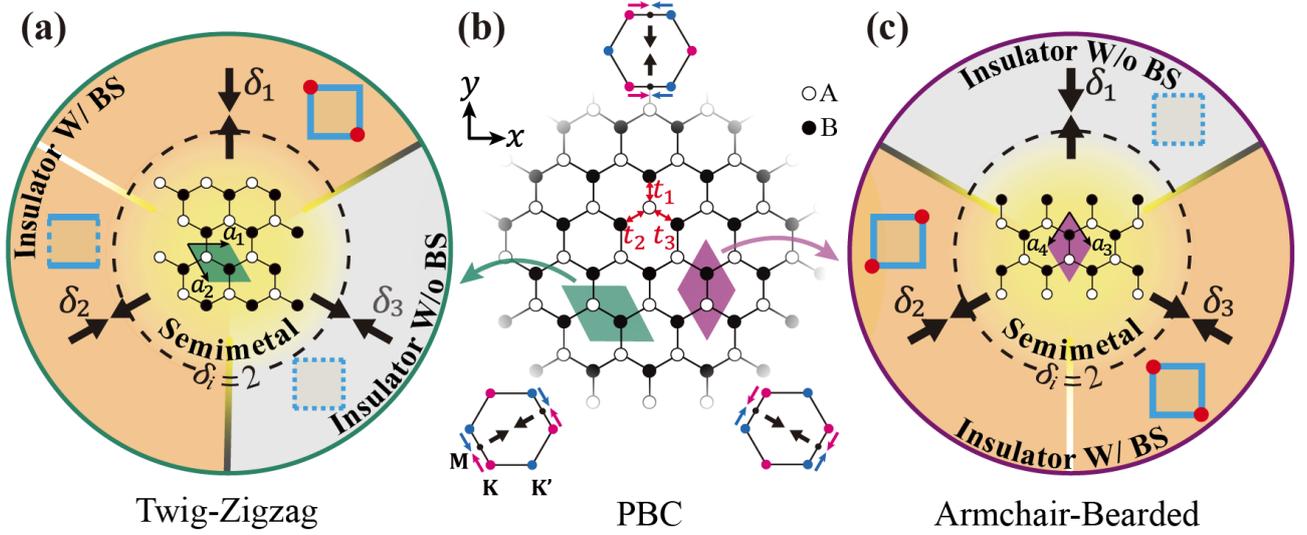

**Fig. 1 Creation and destruction of topological boundary states in uniaxially strained graphene with custom boundaries.** (a) Phase diagram of twig-zigzag flake under uniaxial strain, where the black dashed circle marks the phase transition point $\delta_n = t_n/t_0 = 2, n = 1,2,3$. The HCL structure with custom twig-zigzag boundary condition is displayed in the center, where the green rhombus marks the unit cell and $\mathbf{a_1}, \mathbf{a_2}$ are the basic vectors. Black arrows in each fan-shaped section denote three applied strain (compression) directions along $t_i$-coupling direction, leading to a phase transition from the semimetal (center yellow region) to trivial/nontrivial insulating regimes with (W) or without (W/o) boundary states (gray/orange regions). For illustration convenience, these three strain scenarios are depicted in the same graphene, but it should be understood that they occur independently rather than simultaneously. The square decorated with blue solid (dashed) lines in each fan-shaped section indicates the presence (absence) of edge states in insulating regime, and red dots indicate the corner states. (b) The HCL structure with $A$ (white dots) and $B$ (black dots) sublattices under periodic boundary condition (PBC). Symbols $t_n$ ($n = 1, 2, 3$) represent the three nearest-neighbor couplings. Three hexagons at the corners depict the first 2D Brillouin zone of the HCL, with black arrows inside indicating the directions of applied uniaxial strain. Colored dots and arrows along the edges of each hexagon mark the positions of the Dirac points and their corresponding directions of movement due to the strain, respectively. (c) Phase diagram with the same layout as (a) but under the armchair-bearded boundary condition; the unit cell (purple rhombus) is shown at the center.

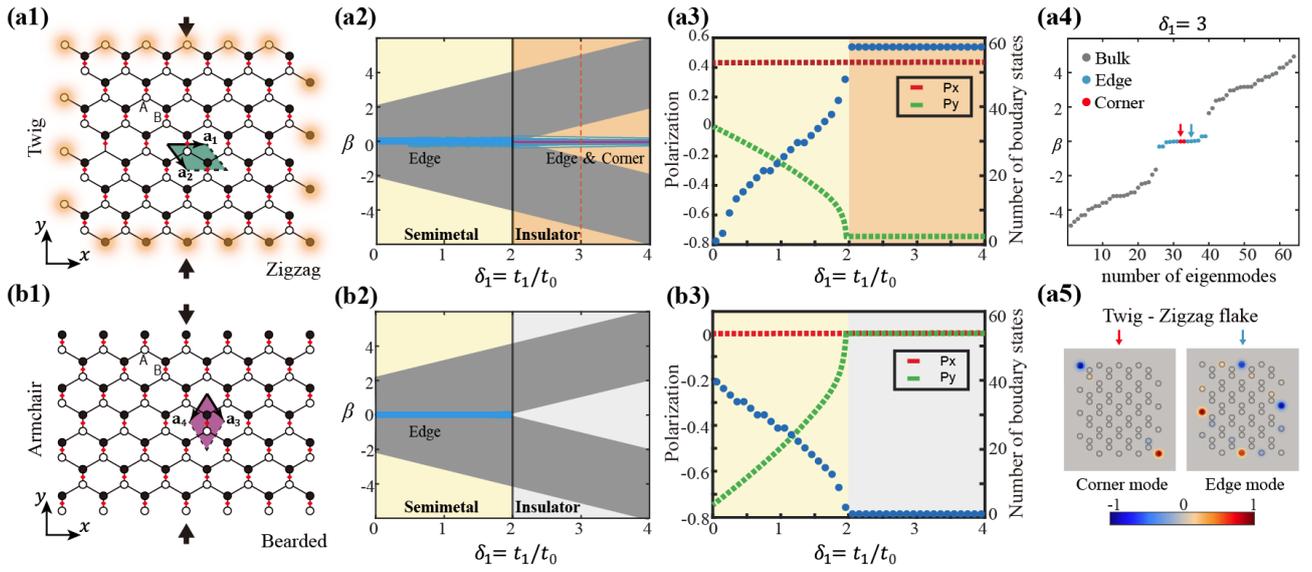

**Fig. 2 Topological boundary states and phase transition in vertically strained graphene flakes.** (a1) The twig-zigzag flake in the nontrivial insulating regime, with the Wannier centers (red markers) located at the edges of each unit cell. The bright orange dots illustrate the presence of boundary states. (a2) The eigenvalue spectrum $\beta$ of the twig-zigzag flake as a function of $\delta_1$, where the eigenvalues of corner and edge modes are highlighted by red and blue lines, respectively. The vertical solid black line indicates the transition point between the semimetal (yellow area) and insulating regimes (orange area). (a3) The evolution of the polarization (red and green dashed lines) and the number of boundary states (blue circles) with respect to $\delta_1$. (a4) The eigenvalue distribution of (a1) under $\delta_1 = 3$ (dashed orange line in (a2)), where the dots pointed by arrows correspond to the mode distributions of corner and compact edge states shown in (a5). (b1), (b2) and (b3) have the same layout as (a1), (a2) and (a3), but for the armchair-bearded flake. The system enters the trivial insulating regime (gray area in (b2) and (b3)) when $\delta_1 > 2$, where no edge or corner states are found.

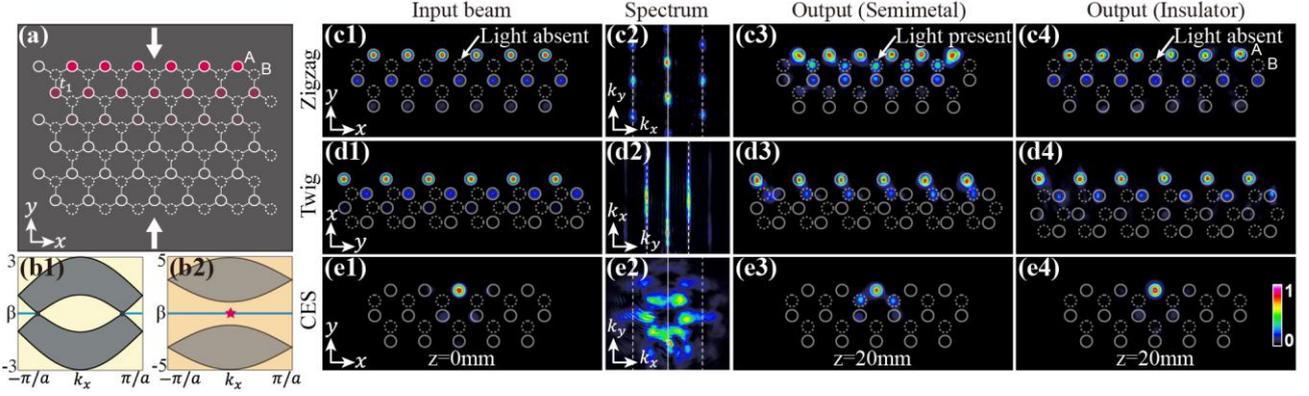

**Fig. 3 Experimental demonstration of extended and compact edge states in strained photonic lattice.** (a) Illustration of HCL with the zigzag boundary condition along the x-direction. The red dots indicate the distribution of zigzag edge state at the $\Gamma$ point in the nontrivial insulating regime (red star in (b2)). The white arrows indicate the direction of uniaxial strain. (b) 1D band structure of HCL with zigzag edge in the semimetal (b1) and insulating (b2) regimes, where the blue lines represent the regions of edge states. (c1) Intensity distribution of the input beam, matching the eigenmode of zigzag edge states at the $\Gamma$ point. (c2) The corresponding Fourier spectrum of (c1), where white lines mark the $\Gamma$ point in the 1D Brillouin zone. (c3, c4) Output of the probe beam in the semimetal ($\delta = 1$) (c3) and the insulating ($\delta > 2$) (c4) regimes. (d) Panels have the same layout as (c), but under the twig boundary condition. (e) The compact edge state excitation along the zigzag boundary. (e1) Intensity distribution of the probe beam at the input, and (e2) the corresponding Fourier spectrum. (e3, e4) Output after 20 $mm$ propagation in the semimetal (e3) and insulating (e4) regimes.

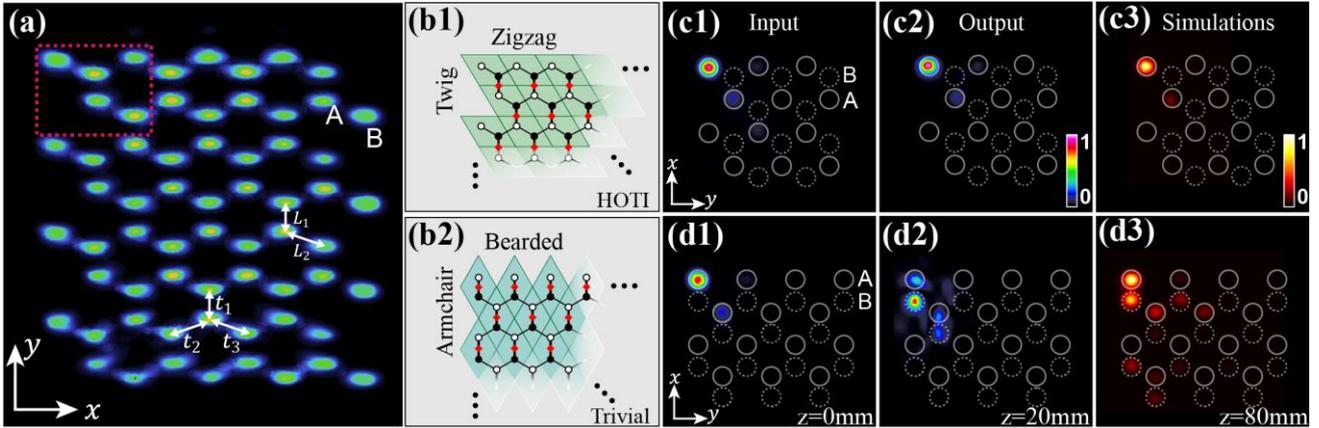

**Fig. 4 Experimental demonstration of corner states in strained photonic graphene in the HOTI regime.** (a) A strained photonic HCL flake with custom zigzag and twig boundaries established in the experiment, where the dashed red rectangle marks the corner excitation position of the probe beam in (c1). $A$ and $B$ mark the two sublattices. The distances between nearest-neighbor sites are $L_1 = 29\ \mu m$ and $L_2 = 40\ \mu m$, corresponding to a nontrivial insulating phase. (b) Schematic diagrams of the corner of the HOTI (b1) and the trivial insulator (b2), where the rhombuses represent the corresponding unit cells compatible with the boundary conditions, and red markers represent the Wannier centers. (c) The corner mode excitation under the twig-zigzag boundary condition. Intensity distribution of the probe beam at input (c1), output ($Z = 20\ mm$) from the experiment (c2), and output ($Z = 80\ mm$) from the simulation (c3). (d) Panels have the same layout as (c), but for the corner excitation in the trivial regime under the armchair-bearded boundary condition, showing the absence of corner states as light populates both sublattices and dissipates strongly into the bulk.